\documentclass[10pt,conference]{IEEEtran}

\usepackage[noadjust]{cite}

\usepackage{array,colortbl,multirow,multicol,hhline,booktabs,ctable}

\usepackage[ruled,vlined,lined,linesnumbered]{algorithm2e}
\let\oldnl\nl
\newcommand{\nonl}{\renewcommand{\nl}{\let\nl\oldnl}}

\ifCLASSINFOpdf
\else
\fi
\usepackage{amssymb}
\usepackage[cmex10]{amsmath}
\usepackage{algorithmic}
\usepackage{array}
\usepackage{mdwmath}
\usepackage{mdwtab}
\usepackage[tight,footnotesize]{subfigure}
\usepackage{url}
\begin{document}
\title{Partitioning Algorithms for Improving Efficiency of \\Topic Modeling Parallelization}

\author{
\IEEEauthorblockN{Hung Nghiep Tran}
\IEEEauthorblockA{University of Information Technology\\VNU-HCMC\\Vietnam\\Email: nghiepth@uit.edu.vn}
\and
\IEEEauthorblockN{Atsuhiro Takasu}
\IEEEauthorblockA{National Institute of Informatics\\SOKENDAI (The Graduate University for Advanced Studies)\\Japan\\Email: takasu@nii.ac.jp}
}

\maketitle

\begin{abstract}
Topic modeling is a very powerful technique in data analysis and data mining but it is generally slow. Many parallelization approaches have been proposed to speed up the learning process. However, they are usually not very efficient because of the many kinds of overhead, especially the load-balancing problem. We address this problem by proposing three partitioning algorithms, which either run more quickly or achieve better load balance than current partitioning algorithms. These algorithms can easily be extended to improve parallelization efficiency on other topic models similar to LDA, e.g., Bag of Timestamps, which is an extension of LDA with time information. We evaluate these algorithms on two popular datasets, NIPS and NYTimes. We also build a dataset containing over 1,000,000 scientific publications in the computer science domain from 1951 to 2010 to experiment with Bag of Timestamps parallelization, which we design to demonstrate the proposed algorithms' extensibility. The results strongly confirm the advantages of these algorithms.
\end{abstract}

\IEEEpeerreviewmaketitle

\section{Introduction}
Topic modeling is a very powerful technique in data analysis and data mining. Thus, it has been used quite frequently in many research areas. However, these models, from the first and most simple models such as LDA \cite{blei2003latent} to other more advanced models such as Bag of Timestamps (BoT) \cite{masada2009bag}, have one significant drawback in practice: they are generally slow.

A common solution for this problem is parallelization of model learning. One of the first parallel algorithms was AD-LDA by Newman et al., which makes multiple data copies to sample in multiple processes, then synchronizes them \cite{newman2007distributed}. This approach is simple but it requires large memory and a costly synchronization process after each sampling iteration. 

A more sophisticated algorithm was proposed by Yan et al. based on partitioning data \cite{yan2009parallel}. This algorithm theoretically guarantees that speedup is near linear, because the overhead is theoretically very small. Moreover, space complexity is almost constant with respect to the number of parallel processes. However, this algorithm is not very popular because it must deal with the problem of load balancing, which makes it less useful in practice.

The above algorithm partitions both documents and words to divide the document--word matrix into $P \times P$ parts, where $P$ is the number of parallel processes. With this partitioning scheme, partitions on the main diagonal line and on each parallel diagonal line are nonoverlapping with respect to documents and words, so these partitions are read--write nonconflicting and could be sampled in parallel on shared data. Hence, this algorithm reduces memory consumption and synchronization cost compared with algorithms similar to AD-LDA.

However, when running in parallel, the slowest process must finish before the next sampling iteration can start, so all the other processes must wait. Load balancing is the major overhead that reduces parallelization efficiency of this algorithm. Thus, to increase parallelization efficiency, work should be distributed evenly among these processes.

The partitioning scheme is the main point of this algorithm but it also makes load balancing very difficult because it is not trivial to divide the document--word matrix so that each partition contains the same number of word tokens. The exact solution of this problem is equivalent to an NP-hard integer programming problem. Current partitioning algorithms are naive randomized algorithms that must run for a long time but load balancing is still low.

We address this problem by developing three partitioning algorithms based on heuristics to distribute word tokens evenly in the document--word matrix. The first two are simple deterministic algorithms. They give good load balancing and run much more quickly than current randomized algorithms. The third algorithm is a randomized algorithm based on some sophisticated conditions to permute the document--word matrix. This algorithm requires a similar run time as current randomized algorithms but it gives much better load balancing.

Our partitioning algorithms are also extensible to other models similar to LDA, e.g., BoT, which is an extension of LDA allowing topic modeling with time information. Currently, BoT does not have a parallel algorithm, so we designed a parallel algorithm for BoT based on Yan et al.'s parallel LDA and apply our partitioning algorithms to improve the parallelization efficiency of this model.

We evaluated the proposed partitioning algorithms using two popular datasets, NIPS and NYTimes, to demonstrate the load-balancing improvement. We also built a dataset containing over 1,000,000 scientific publications in the computer science domain from 1951 to 2010 with information about published year. We use this dataset to evaluate the designed BoT parallelization.

This research presents the following main contributions.
\begin{enumerate}
  \item We develop three partitioning algorithms to address the load-balancing problem in parallelization of topic modeling. These algorithms either run more quickly or achieve better load balancing than current partitioning algorithms.
  \item To demonstrate the extensibility of the proposed algorithms, we design a parallel algorithm for BoT and apply the proposed partitioning algorithms to improve its parallelization efficiency.
  \item We build and publish a dataset containing over 1,000,000 scientific publications in the computer science domain with information about published year. This dataset could be used to experiment with time-aware topic models. We demonstrate analysis of this dataset using the designed BoT parallelization.
\end{enumerate}

Section 2 presents a summary of related research. In Section 3, we define the problem of load balancing and Section 4 presents the proposed approach. Our experiments are described in Section 5, and results and discussion follow in Section 6. Section 7 concludes.
\section{Related Work}
Topic modeling is a powerful technique in text analysis and data mining. One of the first models was LDA developed by Blei et al. in 2003 \cite{blei2003latent}. Since then, many more advanced models have been proposed, especially models that incorporate time information like Dynamic Topic Model \cite{blei2006dynamic} and Topic over Time model \cite{wang2006topics}.

Topic modeling algorithms have one significant drawback: they are generally quite slow. With the appearance of more advanced algorithms, this problem has become more important. Many solutions to this problem have been proposed; most of them use parallelization. 

One of the first studies of topic modeling parallelization was AD-LDA by Newman et al. \cite{newman2007distributed}. This algorithm is very simple but it requires multiple copies of the data and a costly synchronization process after each sampling iteration. Many other parallel algorithms have been proposed. These algorithms could be categorized into three main approaches: (1) \textit{Copy and Sync}, like AD-LDA \cite{newman2007distributed,smyth2009asynchronous,masada2009accelerating}, (2) \textit{Nonblocking Algorithms}, which use atomic operations to use shared data \cite{smola2010architecture}, and (3) \textit{Data Partitioning}, like the algorithm of Yan et al. \cite{yan2009parallel}.

Data partitioning-based algorithms are a very promising approach because they theoretically guarantee a near-linear speedup and do not require extra space for data copies. However, these algorithms have one significant drawback: it is difficult to achieve load balancing \cite{yan2009parallel}. In practice, therefore, the speedup is usually not as good as expected. 

Few studies have investigated this load-balancing problem. Current partitioning algorithms are naive randomized algorithms that do not give good load balancing \cite{yan2009parallel}. In this research, we address this load-balancing problem.

\section{Problem Definition}
\subsection{Parallel algorithm and partitioning scheme}
Many parallelizations have been proposed for LDA algorithms, especially collapsed Gibbs sampling \cite{griffiths2004finding, newman2007distributed, newman2009distributed, smyth2009asynchronous, masada2009accelerating, smola2010architecture}. As noted, we consider the parallel algorithm proposed by Yan et al. in 2009 \cite{yan2009parallel}. This algorithm theoretically guarantees that speedup is near linear, i.e., the overhead is very small. Moreover, space complexity is almost constant with respect to the number of parallel processes.

\begin{figure}[h]
\centering
\includegraphics[width=0.25\linewidth]{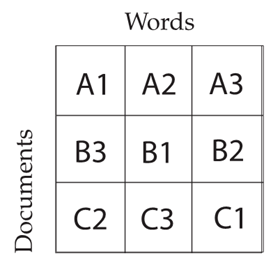}
\caption{Partitions of a document--word matrix.}
\label{fig:Partition}
\end{figure}

The main point of this algorithm is the partitioning scheme. Unlike AD-LDA, which only partitions documents into many parts, this algorithm partitions both documents and words. The document--word matrix is divided both horizontally and vertically into $P \times P$ parts, where $P$ is the number of processes in parallel. For example, Figure \ref{fig:Partition} demonstrates a document--word matrix, which is partitioned $3 \times 3$ ways \cite{ihler2012understanding}. The main diagonal line contains partitions \textit{A1, B1, C1}, other parallel diagonal lines contain partitions \textit{A2, B2, C2} and \textit{A3, B3, C3}, respectively.

With this partitioning scheme, partitions on the main diagonal line and other parallel lines are nonoverlapping with respect to documents and words. Thus, parallel collapsed Gibbs sampling on these partitions is read--write nonconflicting on the document--word counting matrix and the topic--word counting matrix. As a result, this algorithm only requires one copy of each of these two matrices.

However, this partitioning scheme also requires that sampling on each diagonal line must be finished completely before sampling on the next diagonal line. Thus, when running in parallel, all other processes must wait for the slowest process to finish. To improve the efficiency of this algorithm, load balancing between parallel processes must be improved.

\subsection{Load-balancing problem}
\label{subsecteta}
In collapsed Gibbs sampling, the basic operation is topic sampling for a word token. For load balancing, work should be distributed evenly among all processes. Thus, all partitions that are processed in parallel should contain the same number of word tokens.

With the given partitioning scheme, it is not easy to achieve load balance. The current partitioning algorithm proposed by Yan et al. provides poor load balance and must be run for a long time for better results \cite{yan2009parallel}.

To present our algorithms, we formalize the problem in a similar way to that used by Yan et al. \cite{yan2009parallel}. We define $J$ as the set of documents and $V$ as the set of words. The number of documents is $D$, the number of words is $W$, the number of word tokens is $N$. The partitioning algorithm divides $J$ into $P$ disjoint subsets $J_1, \ldots, J_P$ and $V$ into $P$ disjoint subsets $V_1, \ldots, V_P$. Then the document--word matrix $DW$ is divided into $P \times P$ corresponding partitions $DW_{mn}, m \in \{1, \ldots, P\}, n \in \{1, \ldots, P\}$. For $n \in \{1, \ldots, P\}$ and $m \oplus n = (m + n)$ mod $P$, $DW_{m, m \oplus n}$ form diagonal lines having partitions that are sampled in parallel by $P$ processes.

We define the workload matrix $R = (r_{jw})$, where $r_{jw}$ is the number of occurrences of word $w$ in document $j$. $RR$ is a list of rows and $CR$ is a list of columns of $R$. We define the submatrix $R_{mn} = (r_{jw}) \forall j \in J_m, w \in V_n$ corresponding to the partition $DW_{mn}$. We define the cost of this partition as $C_{mn} = \sum_{r_{jw} \in R_{mn}} r_{jw}$. Because all other processes must wait for the slowest process to finish, the cost of each parallel epoch is the maximum cost of the parallel partitions. Thus, the optimal data partition is equivalent to minimizing the following cost function:
\begin{equation}
C = \sum_{l = 0}^{P - 1} \max_{(m,n):m \oplus l = n}{C_{mn}}.
\end{equation}
We define the optimum cost $C_{opt} = \sum_{j \in J, w \in V} r_{jw}/P$. Then, we define the load-balancing ratio $\eta$ as 
\begin{equation}
\eta = C_{opt} / C.
\end{equation}

Exact optimization of $\eta$ is equivalent to an NP-hard integer programming problem \cite{boyd2004convex}. Thus, instead of an exact solution, we use other approaches to optimize $\eta$.

We define the row workload or the length of row $RR_j = \sum_{w \in V}{r_{jw}}$. In case we partition the document--word matrix, this is the number of word tokens in document $j$. Similarly, we define the column workload or the length of column $CR_w = \sum_{j \in J}{r_{jw}}$. In case we partition the document--word matrix, this is the number of word tokens of word $w$ in every document.

We can compare the lengths of rows or columns to identify longer and shorter ones. Then we can sort rows and columns based on their length and find the \textit{longest row}, \textit{longest column}, \textit{shortest row}, and \textit{shortest column}.

To achieve the optimum partitions, matrix $R$ should be distributed evenly. We propose algorithms to distribute workload evenly and divide $R$ into partitions with equal workload.

\section{Proposed Approach}
First, we propose three heuristics for permuting a matrix to create an evenly distributed matrix. Then, we develop three algorithms to divide $R$ into $P \times P$ approximately equal parts.

\subsection{Heuristics}
\textbf{Heuristic 1.} To make a matrix more evenly distributed, interpose a long row and a short row from the beginning of the row list and interpose a long column and a short column from the beginning of the column list.

For example, we could make $RR_1$ the \textit{longest row}, $RR_2$ the \textit{shortest row}, $RR_3$ the \textit{second longest row}, $RR_4$ the \textit{second shortest row}, $\ldots$, $RR_{D}$ the \textit{medium length row}. Similarly for the columns, we could make $CR_1$ the \textit{longest column}, $CR_2$ the \textit{shortest column}, $CR_3$ the \textit{second longest column}, $CR_4$ the \textit{second shortest column}, $\ldots$, $CR_{W}$ the \textit{medium length column}.

\textbf{Heuristic 2.} To make a matrix more evenly distributed, interpose a long row and a short row from both the beginning and the end of the row list and interpose a long column and a short column from both the beginning and the end of the column list. 

For example, we could make $RR_1$ the \textit{longest row}, $RR_2$ the \textit{shortest row}, $RR_{D}$ the \textit{second longest row}, $RR_{D-1}$ the \textit{second shortest row}, $\ldots$, $RR_{D/2}$ the \textit{medium length row}. Similarly, we could make $CR_1$ the \textit{longest column}, $CR_2$ the \textit{shortest column}, $CR_{W}$ the \textit{second longest column}, $CR_{W-1}$ the \textit{second shortest column}, $\ldots$, $CR_{W/2}$ the \textit{medium length column}.

\textbf{Heuristic 3.} This is a generalization of Heuristics 1 and 2. To make a matrix more evenly distributed, interpose rows with different lengths and columns with different lengths, i.e., for every range on the row list and the column list, there should be rows and columns with all kinds of lengths: from longest, medium, to shortest. This heuristic tries to distribute row lengths and column lengths evenly on the row list and the column list, respectively.

Please note that the considered matrix is not symmetric, so other similar permutations can be achieved by swapping the resulting matrix symmetrically vertically and/or horizontally after applying these heuristics.

\subsection{Partitioning algorithms}
Heuristics 1 and 2 are simple, so they can be used directly to develop the algorithms. Heuristic 3 is more complicated. It is used as an inspiration to develop the third algorithm, which is more sophisticated. The first two algorithms are deterministic and the third one is randomized.

\textbf{Algorithm 1.} This algorithm permutes rows and columns based on \textbf{Heuristic 1}, then divides rows into $P$ parts with approximately equal numbers of word tokens, and similarly for columns.

\begin{algorithm} [h]
\SetKwInOut{Input}{Input}
\SetKwInOut{Output}{Output}

\Input{Matrix workload $R$}
\Output{Partitions of documents $J_1, \ldots, J_P$ and partitions of words $V_1, \ldots, V_P$}

\nonl// Permute rows.

Sort the row list $RR$ in descending order.

\For {each row $RR_i$ in $RR$} {
	\If{i mod 2 = 0} {
		Insert the last row $RR_D$ before $RR_i$.
		
		Remove the last row.
	}
}
	
\nonl// Permute columns.

Sort the column list $CR$ in descending order.

\For {each column $CR_i$ in $CR$} {
	\If{i mod 2 = 0} {
		Insert the last column $CR_D$ before $CR_i$.
		
		Remove the last column.
	}
}

\nonl// Partition rows and columns

Divide $RR$ into $P$ consecutive groups $J_1, \ldots, J_P$, each one having an equal number of word tokens.

Divide $CR$ into $P$ consecutive groups $V_1, \ldots, V_P$, each one having an equal number of word tokens.

\caption{Data partitioning Algorithm \textit{A1}.}	
\label{A1}
\end{algorithm}

\textbf{Algorithm 2.} Similar to \textbf{Algorithm 1}, but this algorithm permutes rows and columns based on \textbf{Heuristic 2}.

\begin{algorithm} [h]
\SetKwInOut{Input}{Input}
\SetKwInOut{Output}{Output}

\Input{Matrix workload $R$}
\Output{Partitions of documents $J_1, \ldots, J_P$ and partitions of words $V_1, \ldots, V_P$}

\nonl// Permute rows.

Sort the row list $RR$ in descending order.

\For {each row $RR_i$ in $RR$, $i < D/2$} {
	\If{i mod 2 = 0} {
		Swap row $RR_i$ with row $RR_{D+1-i}$.
	}
}
	
\nonl// Permute columns.

Sort the column list $CR$ in descending order.

\For {each column $CR_i$ in $CR$, $i < V/2$} {
	\If{i mod 2 = 0} {
		Swap column $CR_i$ with column $CR_{V+1-i}$.
	}
}

\nonl// Partition rows and columns

Divide $RR$ into $P$ consecutive groups $J_1, \ldots, J_P$, each one having an equal number of word tokens.

Divide $CR$ into $P$ consecutive groups $V_1, \ldots, V_P$, each one having an equal number of word tokens.

\caption{Data partitioning Algorithm \textit{A2}.}
\label{A2}
\end{algorithm}

\textbf{Algorithm 3.} Given the number of parallel processes $P$, we build $P$ ranges on the row list. Each of them has rows with all kinds of length and similarly for columns. To do this, the algorithm randomly shuffles workload matrix $R$ under the restrictions that row and column lengths are distributed evenly. With these restrictions, this algorithm is guaranteed to achieve better load balance than simply shuffling rows and columns as in Yan et al.'s algorithm.

Then we divide the rows of matrix $R$ into $P$ parts with approximately equal numbers of word tokens and similarly for columns. This is a randomized algorithm, so we repeat the process several times and compute the load-balancing ratio $\eta$ each time; then we find the partition set with the highest $\eta$. This algorithm runs quite quickly, because it contains only single loops. Moreover, every loop could be executed in parallel. In practice, this algorithm has approximately the same running time as the partitioning algorithm proposed by Yan et al.

\begin{algorithm} [h]
\SetKwInOut{Input}{Input}
\SetKwInOut{Output}{Output}

\Input{Matrix workload $R$}
\Output{Partitions of documents $J_1, \ldots, J_P$ and partitions of words $V_1, \ldots, V_P$}

Create empty temporary lists $RT = {RT_1, \ldots, RT_P}$, and $CT = {CT_1, \ldots, CT_P}$.

\nonl// Permute rows.

Sort the row list $RR$ in descending order.

Divide items in $RR$ into groups of $P$ consecutive items.

\For {each group of $P$ consecutive items in $RR$} {
	Uniformly randomly shuffle items in this group.
	
	Assign each item $i$ in this group to the list $RT_i$.
}
	
Set the row list $RR$ to empty.

\For {each temporary list $RT_i$ in ${RT_1, \ldots, RT_P}$} {
	Uniformly randomly shuffle items in $RT_i$.
	
	Append all items in $RT_i$ to the end of $RR$.
}

\nonl// Permute columns.

Sort the column list $CR$ in descending order.

Divide items in $CR$ into groups of $P$ consecutive items.

\For {each group of $P$ consecutive items in $CR$} {
	Uniformly randomly shuffle items in this group.
	
	Assign each item $i$ in this group to the list $CT_i$.
}
	
Set the column list $CR$ to empty.

\For {each temporary list $CT_i$ in ${CT_1, \ldots, CT_P}$} {
	Uniformly randomly shuffle items in $CT_i$.
	
	Append all items in $CT_i$ to the end of $CR$.
}

\nonl// Partition rows and columns

Divide $RR$ into $P$ consecutive groups $J_1, \ldots, J_P$ having equal numbers of word tokens.

Divide $CR$ into $P$ consecutive groups $V_1, \ldots, V_P$ having equal numbers of word tokens.

\caption{Data partitioning Algorithm \textit{A3}.}	
\label{A3}
\end{algorithm}

Algorithms \textit{A1} and \textit{A2} are deterministic so they require only one run time, in contrast to the randomized algorithm, which require tens or even hundreds of run times to achieve good results. Hence, \textit{A1} and \textit{A2} could be as much as two orders of magnitude faster than the randomized algorithm.

\subsection{Parallel algorithm for BoT}
\begin{figure}[h]
\centering
\includegraphics[width=0.375\linewidth]{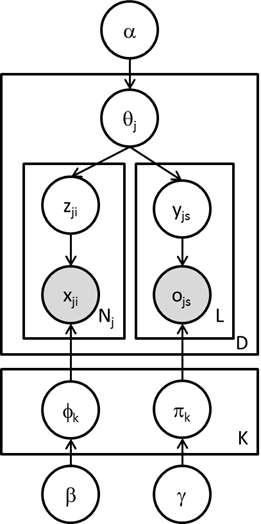}
\caption{Bag of Timestamps graphical model.}
\label{fig:BoT}
\end{figure}

BoT is an extension of LDA that makes use of time information, e.g., a paper's year of publication. It is similar to Dynamic Topic Model \cite{blei2006dynamic} and Topic over Time model \cite{wang2006topics}. Figure \ref{fig:BoT} presents the graphical model of BoT with a similar notation to that for LDA. In this model, each document $J_j, j = 1 \ldots D$ is attached to a timestamp array $TS_j = \{o_{js}, s = 1, \ldots, L\}$ with length $L$. This array is considered to extend the content of document $J_j$ and timestamps are considered equivalent to words. Each timestamp $ o_{js} $ has the topic assignment $ y_{js} $. Timestamps share the topic per document distribution $ \theta $ with words, but also have their own timestamps per topic distribution $ \pi $ with prior $ \gamma $ \cite{masada2009bag}.

BoT gives us information about the presence of a topic in the time line among other advantages over LDA \cite{masada2009bag}. However, collapsed Gibbs sampling for BoT must sample timestamps in addition to words in each document, which increases the learning time and makes BoT less popular in practice. To exploit the benefits of BoT, we design a parallel algorithm for learning BoT and demonstrate that our partitioning algorithms are easily extensible to models similar to LDA.

In BoT, there are two topic assignment matrices, the document--word matrix $DW$ and the document--timestamp matrix $DTS$. The document--topic counting matrix $C_{Theta}$ is updated based on both $DW$ and $DTS$. The topic--word counting matrix $C_{Phi}$ is updated based on $DW$ and the topic--timestamp counting matrix $C_{Pi}$ is updated based on $DTS$.

For parallel BoT, we first partition $DW$ into $P \times P$ partitions as in Yan et al.'s parallel algorithm. Then, we also partition $DTS$ into $P \times P$ partitions, so partitions on the diagonal are nonconflicting with respect to documents and timestamps. Another approach is to merge the timestamp array into the document content, then partition and sample both words and timestamps in one matrix. In this research, we use the first approach for its simplicity.

With this partitioning scheme, each sampling iteration of the whole $DW$ and $DTS$ matrices requires $P$ epochs. In each one, we do parallel processing on partitions in a diagonal line. In each epoch, we consecutively sample all words in a diagonal line of $DW$, then all timestamp--topic assignments in a corresponding diagonal line of $DTS$.

Similar to the parallel algorithm for LDA, load balancing is a critical problem in this algorithm. Our partitioning algorithm is easy to apply to BoT. We treat the matrices $DTS$ and $DW$ equivalently, so we define a workload matrix $R'$ corresponding to $R$. In $R'$, rows still correspond to documents but columns correspond to timestamps. With these differences noted, we apply the same partitioning algorithm to $R'$ and obtain $P \times P$ partitions for the $DTS$ matrix as $J'_1, \ldots, J'_P$ and $V'_1, \ldots, V'_P$.

\section{Experimental Settings}
\subsection{Datasets}
For experiments on our partitioning algorithms for LDA, we used the two popular datasets NIPS and NYTimes\footnote{\url{http://archive.ics.uci.edu/ml/datasets/Bag+of+Words}}. These datasets were preprocessed by removing stop words but not stemming.

To experiment with BoT, we needed a dataset with time information. We therefore built a dataset from Microsoft Academic Research\footnote{\url{http://academic.research.microsoft.com}} (MAS), which contains over 1,000,000 scientific publications in the computer science domain from 1951 to 2010. Each document has a title and an abstract. We removed stop words and applied the Lovins stemmer using Weka\footnote{\url{http://www.cs.waikato.ac.nz/ml/weka}} \cite{lovins1968development}. Table \ref{tab:Dataset} shows the statistics of our datasets.

\begin{table}[htbp]
  \centering
  \caption{Datasets.}
    \begin{tabular}{r|r|r|r}
    \hline
    Dataset & NIPS & NYTimes & MAS \\
    \hline
    Documents, $D$ & 1500 & 300,000 & 1,182,744 \\
    Unique words (Not Stemmed), $W$ & 12,419 & 102,660 & 728,546 \\
    Unique words (Stemmed), $W$ & N/A   & N/A   & 402,252 \\
    Word instances, $N$ & 1,932,365 & 99,542,125 & 92,531,014 \\
    Unique timestamps, $WTS$ & N/A   & N/A   & 60 \\
    First timestamp & N/A   & N/A   & 1951 \\
    Last timestamp & N/A   & N/A   & 2010 \\
	\hline
    \end{tabular}%
  \label{tab:Dataset}%
\end{table}%

\subsection{Measurement metrics}
To measure the efficiency of partitioning algorithms, we computed the load-balancing ratio $\eta$ as defined in Section \ref{subsecteta}. To our knowledge, the algorithm proposed by Yan et al. is currently the best algorithm \cite{yan2009parallel}. We used this algorithm as the baseline to evaluate our algorithms.

We also wanted to confirm that parallelization does not affect the quality of extracted topics. To do this, we compare the perplexity computed from the nonparallel and parallel algorithms. Lower perplexity indicates that the model has better ability to describe data. For simplicity, we use training set perplexity defined as:
\begin{equation}
Perp(x)=\exp(-\frac{1}{N}\log p(x)),
\end{equation}
with
\begin{equation}
\log p(x) = \sum_{ji}\log\sum_k\theta_{k|j}\phi_{x_{ji}|k}.
\end{equation}

\subsection{Model parameters}
For both LDA and BoT, we set \textit{Number of topics = 256}. The hyperparameters were selected based on the suggestions in the LDA and BoT papers \cite{blei2003latent, masada2009bag}. For LDA, $\alpha = 0.5$, $\beta = 0.1$. For BoT, $\gamma = 0.1$. The length of the timestamp array was set to $L = 16$. These settings are not critical in our experiments.

We ran each model until it converged. As we observed, this required no more than 200 sampling iterations for the burn-in period for all three datasets.

We repeated algorithm \textit{A3} 100 times on NIPS and NYTimes. On MAS dataset, we repeated algorithm \textit{A3} 100 times for $R$ matrix and 200 times for $R'$ matrix.

Our experimental program was developed based on the Java source code for nonparallel collapsed Gibbs sampling LDA provided by Phan et al. \cite{phan2008learning}.

\section{Results and Discussion}
\subsection{Partitioning algorithms }
We show the load-balancing ratios $\eta$ achieved by the proposed algorithms and by the baseline algorithm. Table \ref{tab:nips} shows results for the NIPS dataset. The three proposed algorithms all gave better results than the baseline algorithm. Algorithm \textit{A3} gave the highest $\eta$ and the two deterministic algorithms \textit{A1} and \textit{A2} gave quite good results.

\begin{table}[htbp]
  \centering
  \caption{Load-balancing ratio for NIPS.}
    \begin{tabular}{r|r|r|r|r}
    \hline
    $P$ & \textbf{1} & \textbf{10} & \textbf{30} & \textbf{60} \\
    \hline
    Baseline algorithm & 1.0     & 0.9500 & 0.7800 & 0.5700 \\
    Algorithm A1 & 1.0     & 0.9613 & 0.8657 & 0.7126 \\
    Algorithm A2 & 1.0     & 0.9633 & 0.8568 & 0.7097 \\
    Algorithm A3 & 1.0     & 0.9800 & 0.8929 & 0.7553 \\
    \hline
    \end{tabular}%
  \label{tab:nips}%
\end{table}%

Table \ref{tab:nyt} shows results for the NYTimes dataset. Again, algorithm \textit{A3} gave the best results in all experiments. The other two deterministic algorithms \textit{A1} and \textit{A2} gave very competitive results, which are approximate to the baseline algorithm when $P=10$ and $P=30$ but are higher when $P=60$.

\begin{table}[htbp]
  \centering
  \caption{Load-balancing ratio on NYTimes.}
    \begin{tabular}{r|r|r|r|r}
    \hline
	$P$ & \textbf{1} & \textbf{10} & \textbf{30} & \textbf{60} \\
    \hline
    Baseline algorithm & 1.0     & 0.9700 & 0.9300 & 0.8500 \\
    Algorithm A1 & 1.0     & 0.9559 & 0.9270 & 0.9011 \\
    Algorithm A2 & 1.0     & 0.9626 & 0.9439 & 0.9175 \\
    Algorithm A3 & 1.0     & 0.9981 & 0.9901 & 0.9757 \\
    \hline
    \end{tabular}%
  \label{tab:nyt}%
\end{table}%

\subsection{Parallel learning BoT}
Here we present the results for our parallel algorithm with BoT. Table \ref{tab:bot} shows the perplexity of nonparallel BoT compared with parallel BoT. The perplexity is approximately the same for all cases, i.e., parallelization does not affect the quality of extracted topics. Interestingly, the parallel algorithm achieves a slightly better perplexity than the nonparallel version. This is similar to results reported for other parallel algorithms \cite{newman2007distributed,ihler2012understanding}.

\begin{table}[htbp]
  \centering
  \caption{Perplexity of BoT for the MAS dataset.}
    \begin{tabular}{l|r|r|r}
    \hline
    Algorithm & Nonparallel & Parallel $P=10$ & Parallel $P=30$ \\
    \hline
    Perplexity & 595.2567 & 595.0593 & 593.9016 \\
    \hline
    \end{tabular}%
  \label{tab:bot}%
\end{table}%

\subsection{Discussion}
For data partitioning-based parallel algorithms, the overheads are mostly because of load-balancing problems. Hence, the speedup factor is approximately $\eta \times P$ \cite{yan2009parallel}. Improving the load-balancing ratio $\eta$ directly improves the speedup factor. When $\eta$ is close to one, these parallel algorithms achieve a near-linear speedup.

As we lacked a dedicated server for the experimental environment, we did not record the exact running time of each algorithm. However, in most cases, the running times of partitioning algorithms are small. Typically, Algorithm \textit{A3}'s running time is two orders of magnitude faster than the model training time. Running times for algorithms \textit{A1} and \textit{A2} are two orders of magnitude faster than those of other randomized algorithms, such as Algorithm \textit{A3} and Yan et al.'s algorithm.

The partitioning algorithms improve the efficiency of the parallel algorithm but do not affect the quality of the extracted topics. The perplexities resulting from parallel algorithms are almost unchanged and even slightly better. This could be explained that parallelization adds stochasticity to the model, similarly to other stochastic algorithms, to give a better result \cite{ihler2012understanding,recht2011hogwild}.

As we can see, the deterministic algorithms \textit{A1} and \textit{A2} are quite good in most cases, and they run much more quickly than the randomized counterparts. They should therefore be used in the first place to find whether the load-balancing ratio is good enough. If better load balancing is required, we would use the randomized algorithm \textit{A3}, which is guaranteed to achieve a higher $\eta$ but requires longer running time.
\section{Conclusion}
In this paper, we have addressed the load-balancing problem in the parallelization of topic modeling. We developed three partitioning algorithms that either run more quickly or achieve better load balance than current algorithms. Two deterministic algorithms give good results and run more quickly than the baseline algorithm by two orders of magnitude. The other algorithm is a randomized algorithm that is guaranteed to give much better results but requires running time similar to the baseline algorithm.

These algorithms can easily be extended to other models similar to LDA. We demonstrate their extensibility by designing a parallel algorithm for BoT and apply our partitioning algorithms to improve its parallelization efficiency.

We tested our ideas on two popular datasets NIPS and NYTimes. We also built a dataset containing scientific publications with time information and showed how to analyze this dataset using BoT. The experimental results strongly confirm the advantages of our algorithms.

The proposed algorithms have the potential to improve the applicability of topic modeling, especially advanced models like BoT, on large-scale text data, such as scientific publications. Improving the parallelization efficiency of topic modeling is still an open problem, especially on advanced topic models, and so is an interesting topic for future research.


\bibliographystyle{ieeetran}

\end{document}